# Self-Energy and Excitonic Effects in the Electronic and Optical Properties of TiO$_2$ Crystalline Phases

Letizia Chiodo[1,2], Juan Maria García-Lastra[1], Amilcare Iacomino[3], Stefano Ossicini[4], Jin Zhao[5], Hrvoje Petek[5] and Angel Rubio[1]

[1] Nano-Bio Spectroscopy group and ETSF Scientific Development Centre, Dpto. Física de Materiales, Universidad del País Vasco, Centro de Física de Materiales CSIC-UPV/EHU- MPC and DIPC, Av. Tolosa 72, E-20018 San Sebastián, Spain

[2] IIT Italian Institute of Technology, Center for Biomolecular Nanotechnologies, Via Barsanti, I-73010, Arnesano (Le), Italy

[3] Dipartimento di Fisica "E. Amaldi," Università degli Studi Roma Tre, Via della Vasca Navale 84, I-00146 Roma, Italy, and CNISM, U. di R. Università degli Studi di Napoli "Federico II," Dipartimento di Scienze Fisiche, Complesso Universitario Monte S. Angelo, Via Cintia, I-80126 Napoli, Italy.

[4] Dipartimento di Scienze e Metodi dell'Ingegneria, Università di Modena e Reggio Emilia, Via Amendola 2 Pad. Morselli, I-42100 Reggio Emilia, Italy

[5] Department of Physics and Astronomy, University of Pittsburgh, Pittsburgh, Pennsylvania 15260, USA

**Abstract**

We present a unified ab-initio study of electronic and optical properties of TiO$_2$ rutile and anatase phases, with a combination of Density Functional Theory and Many Body Perturbation Theory techniques. The consistent treatment of exchange-correlation, with the inclusion of many body one-particle and two-particles effects in self-energy and electron-hole interaction, produces a high quality description of electronic and optical properties, giving, for some quantities, the first available estimation for this compound. In particular, we give a quantitative, direct evaluation of the electronic and direct optical gaps, clarifying their role with respect to previous values obtained by various experimental techniques. We obtain a description for both electronic gap and optical spectra that is consistent with experiments, analysing the role of different contributions to the experimental optical gap and relating them to the level of theory used in our calculations. We also show the spatial nature of excitons in the two crystalline phases, highlighting the localization character of different optical transitions. This paper aims at understanding and firmly establishing electro-optical bulk properties, so far not yet clarified, of this material of fundamental and technological interest for green energy applications.

# 1. Introduction

TiO$_2$ is the paradigm for the electronic, optical, and chemical properties of reducible metal oxides. The response of TiO$_2$ single crystal polymorphs and nanostructures to optical excitation is of particular topical interest in solar energy harvesting through photocatalytic splitting of water into hydrogen and oxygen, in photovoltaic generation of electricity, and reduction of CO$_2$ into hydrocarbon fuels.[1-3] The photoactive response of TiO$_2$ nanostructures gained prominence when stable photocatalytic production of H$_2$ upon band gap excitation of TiO$_2$ anatase nanoparticles was demonstrated.[2, 4] The interest in solar energy conversion applications of TiO$_2$ gained further impetus through the invention of dye-sensitized solar cells.[1, 5-6] Reasons for the continued interest in TiO$_2$ for solar energy conversion include the position of its conduction band near the threshold of the energy for hydrogen formation, the relatively long lifetime of conduction band electrons, the high resistance against corrosion, the high earth abundance, and the low cost.[5, 7-8]

Most of the experimental and modelling efforts have been devoted so far to characterization of the surface structure and simple chemical reactions on titanium oxide surfaces[9-20] and nanostructures.[21-23] Incisive theoretical and experimental understanding of the optical properties of TiO$_2$ polymorphs is however clearly essential to advance its suitability for clean energy applications.

In the various technological fields, one of the fundamental quantities determining the overall behaviour of a device is the gap between valence and conduction bands. Depending on the phenomenon, the relevant gap will be the electronic or the optical one. The former one is described by e.g. photoemission experiments, and connected to a one-particle description. The optical gap is instead given by light absorption experiments, and intrinsically characterized by the co-presence of the excited electron and its hole. The electronic band gap is related to e.g. the potentials for the production of H$_2$, or to the electronic charge injection in solar cells. The fraction of the solar spectrum absorbed is connected to the optical gap, which can be tuned by the combination of TiO$_2$ with other solid-state materials, or organic components.[6, 24-30]

Optical properties can be strongly affected by the behaviour of excitons in bulk and nanostructures. Description of excitons, and more in general of optical properties in TiO$_2$ can result not easy, because different aspects have to be taken into account. First of all, the ionic character of the compound results in an important ionic contribution to the screening[31], which is an order of magnitude larger than the pure electronic screening (in rutile, ionic screening values are 111 in the *xy* plane, 257 along *z* [31]). Moreover, a strong exciton-phonon interaction has been observed[31-36] in this compound. The lowest energy excitonic transitions reported for both phases are indeed due to indirect processes, involving phonon contributions. A further complication arises from the presence of self-trapping phenomena.[31-36] The exciton creation induces a local ionic relaxation which traps the exciton, localizing it. Given all these aspects, the nature of excitons, the carrier transport, and the electro-optical role of bulk and surface structure in nanomaterials are still poorly understood. We report in this paper an ab-initio estimation of electronic and optical gaps, and show how the very anisotropic spatial distribution of the first allowed optical transitions is related to the crystal structure of the two phases. These results will be useful when studying the excitonic properties for $TiO_2$ surfaces, interfaces and nanostructures, and when analyzing the effect of defects and local structural deformations on excitons behaviour.

A clear determination of the gap values for the bulk phase material is the first, necessary step for understanding and optimizing processes involving electronic and optical transitions of $TiO_2$ nanostructures. Indeed, despite the clear importance of its surfaces[9-20] and nanostructures[21-23], investigations of $TiO_2$ bulk electronic and optical properties have not provided, so far, a comprehensive description of the material. Most of the experimental and computational work has been focused on the synthesis and the analysis of systems with reduced dimensionality.[9-20 21-23]

In this introduction we aim to review the existing results obtained using a variety of experimental techniques and ab-initio calculations, in order to elucidate the known properties of bulk $TiO_2$. Previous data will be compared, in the body of the paper, with a complete, consistent ab-initio description, which includes many body effects in the calculation of the electronic and optical properties. Most experimental and theoretical data reported from literature refers to the rutile phase (see Fig. 1, top), while anatase (Fig. 1, bottom) in general has been less studied. The anatase polymorph, however, has received recently more attention due in part to its higher photocatalytic activity and greater thermodynamic stability at the nanoscale compared to rutile[4]. The absorption edge has been determined with a good degree of confidence, to be 3.0 eV[37] and 3.2 eV[37] for rutile and anatase, respectively. This edge is determined by the co-presence of various effects: polaronic screening[31], electron-phonon coupling, defects and oxygen vacancies, and is therefore lower in energy than the absorption edge due to direct excitons. While a general agreement seems to exist in literature concerning the optical absorption edge of these materials, values for basic electronic properties such as the electronic band gap still have a large degree of uncertainty. Moreover, the spatial behaviour of excitons in the bulk phase has never been investigated.

Concerning the existing experimental literature on the bulk properties, first measurements on the bulk electronic and optical properties of $TiO_2$ were performed starting from the 1950s [37-49] and have been reinvestigated more recently.[34-35, 50-65]

The electronic properties of valence states of rutile $TiO_2$ were studied experimentally by angle-resolved photoemission spectroscopy.[54] The valence band of $TiO_2$ consists mainly of O 2$p$ states partially hybridized with Ti 3$d$ states. The metal 3$d$ states constitute the conduction band, with a small amount of mixing with O 2$p$ states. From the symmetry of the $TiO_6$ octahedra (see Fig. 1), 3$d$ states are split by the crystal field into low energy $t_{2g}$ and high energy $e_g$ sub-bands. From the photoemission data on rutile[54], it was estimated that the electronic band gap, corresponding to the difference between the valence band maximum (VBM) and the conduction band minimum (CBM), is about 4~eV. This was the observed binding energy of the first states below the Fermi energy and it was in agreement with other reported data from electron energy loss spectroscopy[39] and previous ultraviolet photoemission spectroscopy (UPS).[47, 49, 51] At the same time, combined photoemission and inverse photoemission experiments[56] provided an electronic band gap of 3.3±0.5 eV. The large uncertainty on this measurement arose from the low instrumental resolution. Very recent and more accurate photoemission and inverse photoemission experiments have shown that the electronic band gap of rutile (110) surface is 3.6±0.2 eV eV [65]. Assuming that the bulk electronic gap determination is not strongly affected by the surface states, we can adopt the value of 3.3 ÷ 3.6 eV as a reference for rutile in the following discussion, keeping also in mind the previous data of 4.0 eV.[54] In any case a quite large degree of uncertainity exists on the determination of the electronic rutile gap.

The electronic structure of rutile bulk has also been described using other experimental techniques, such as electrical resistivity[39], electroabsorption[40-41], photoconductivity and photoluminescence[42, 57], X-ray absorption spectroscopy (XAS)[43-44, 46, 48, 50, 53] resonant Raman spectroscopy[52, 57], photoelectrochemical analysis[38, 66] and electron energy loss spectroscopy (EELS)[60, 67-68]. These experiments have provided details of $TiO_2$ electronic properties, in particular concerning the hybridization between Ti 3$d$ and O 2$p$ states and the characterization of valence and conduction bands close to the gap.

A similar kind of experimental description is available for the anatase polymorph. There are several XAS[35, 53, 55] and photoemission[62, 64] measurements concerning its electronic structure, characterizing the *p-d* hybridization of Ti-O bonds. To our knowledge, however, there are no available estimations in the literature of the electronic band gap from combined photoemission and inverse photoemission experiments for anatase.

We proceed now reviewing the experimental results for optical properties of $TiO_2$. Such measurements are of great importance for the photocatalytic and photovoltaic applications of this material. As already said, from the optical absorption spectra of both phases[37] the room temperature absorption edge is found to be 3.0~eV for rutile, and 3.2~eV for anatase.[37] It is important to emphasize here that the experimentally determined optical edge does not correspond to the optical band gap given by the first allowed optical transition (with a certain degree of excitonic character). The optical direct transitions are evaluated within the ab initio methods used here. In the experimental optical edge, higher-order processes, such as absorption of two photons, or the simultaneous absorption of a photon and scattering with a phonon, can be involved. The ionic screening is also present in experimental data. Moreover, the presence of defects, such as oxygen vacancies, can significantly affect optical properties. Both defects and phonons will be present in any experimental sample of the material at finite temperatures. These observations have to be kept in mind when directly comparing experimental measurements with the theoretical results presented in the following.

The absorption edge has been investigated in detail for rutile by absorption, photoluminescence, and Raman scattering techniques.[45] These methods provide description of the optical edge, due to indirect transitions induced by phonon-exciton interactions. The edge at 3.031~eV is associated with a $2p_{xy}$ exciton, while a lower energy $1_s$ quadrupolar exciton has been identified below 3~eV. The first dipole allowed gap is at 4.2~eV[45] and, according to the combined results of these three techniques, an exciton is reported at 3.57 eV.[45]

Concerning anatase[58-59], the optical spectrum has been re-evaluated[59] confirming the absorption edge at 3.2~eV, again due to indirect transitions, as shown from investigation[34] of anatase's spectrum fine details. It has been inferred, from data on the Urbach tail[34], that excitons in anatase are self-trapped in the octahedron coordinating the titanium atom. This behaviour is in contrast to the rutile phase, where excitons are known to be free due to the different packing of rutile's octahedra[34]. We will return to this point when discussing the spatial distribution of optical excitations in the two phases. The first dipole allowed transition in anatase has not been experimentally estimated, to the best of our knowledge.

We would like to recall that the two types of experiments (photoemission, and optical absorption) provide information on two different physical quantities. Photoemission experiments involve a change in the total number of electrons in the material whereas optical absorption experiments do not. The latter involves the creation of an electron-hole pair in the material, with the hole stabilizing the excited electron. For this reason, the theoretical electronic band gaps obtained by means of Density Functional Theory (DFT)[69-70], DFT with hybrid functionals[71-72] or many-body calculations[69-70, 73] should be compared with photoemission experiments, as it has been done by Oshikiri et. al[73]. On the other side, the experimental optical band gap should be compared with simulations of the spectra that include excitonic effects, as has been recently done by Lawler et al. using the Bethe-Salpeter Equation[74] There is a general trend, however, in theoretical studies, to make incongruous comparisons between the theoretical electronic band gaps and the experimental optical gap values.[69-72, 75-77]

Moreover, there is as an intrinsic propensity of DFT to underestimate the electronic band gaps.[78-79] The electronic band gap obtained at DFT level [70, 80-81] is 1.88 eV for rutile (direct, at $\Gamma$), and 2.05 eV[80] (indirect, from X to $\Gamma$) and 2.36 eV [75] (direct, at $\Gamma$) for anatase. When many body corrections are applied, the resulting electronic gap is around 3.8 eV for both phases[70, 81-82]. Such values, when compared to the optical absorption edge, measured from absorption experiments, clearly give improper results. The optical gaps (3.0-3.2 eV) are larger than the DFT ones and always smaller than the electronic ones calculated with many-body corrections. The optical gap in in general lower than the electronic one for a series of reasons: excitonic effects, phonon-assisted indirect transitions, two-photons processes, temperature effects. When the calculated electronic gap, at many body level, is properly compared[73] to photoemission results,[54, 56, 65] as done in the present paper, a quite good agreement can be found.

Concerning the optical spectra of these polymorphs, some calculations[37, 58, 61, 74, 83] have been performed at the Random Phase Approximation (RPA) level, but this description is inappropriate to correctly take into account the optical response of the material, as shown in the results section. The many body description of optics[74] gives an almost perfect agreement between theoretical and experimental spectra. As discussed in the following, in the theoretical description just direct optical transitions are involved, phonon-assisted ones are not taken into account, so a direct comparison between theoretical edge and experimental one has to make this distinction.

To resume, the theoretical investigations presented in the literature on the structural, electronic, and optical properties of $TiO_2$ have been performed sparsely, focusing just on one of the phases, or on some specific property of interest. The use of different methods and levels of theory makes also the results somehow inconsistent. A comprehensive description of properties of both phases in the same theoretical and computational framework is still missing

The main goals of this work, therefore, is to obtain in a unified description from ab-initio calculations the optical and the electronic properties, in particular, the electronic and optical gaps of both rutile and anatase phases, treating the theoretical framework and complexity for all the investigated properties at the same level. We find electronic gaps for the two phases in quite good agreement with experimental data, when available, and give an estimation of direct optical absorption edge.

A novelty is to calculate the optical spectra at a fully ab-initio level, since we used first-principles gap corrections for optics calculations. We will also show the effect of the many body corrections to the quasi-particle energies on the final absorption spectra and the role of screening anisotropies on the optical response. A comprehensive description of excitons in the two phases will be provided through their spectral analysis, their spatial distribution, and the characterization of the electronic states and k points involved in the transitions. This analysis is expected to be interesting in view of the study of excitons behaviour in nanostructures and surfaces.

This work is organized as follow: we present a short section on the theoretical background, the results for electronic properties and optical behaviour, with a detailed description of excitons, and finally the conclusions.

## 2. Theoretical Methods

The state of the art of DFT combined with many body techniques[84] is used in this work.[85] The structural and energetic description of the rutile and anatase $TiO_2$ bulk phases have been obtained by DFT[86-87], a well established tool for the description of ground state properties. Within DFT we calculate the equilibrium structure, the density of states, and the electronic band structure. Ab-initio DFT calculations were carried out using the plane-waves code PWSCF[88]. In order to be consistent with the full ab-initio philosophy of this work, all the results shown here have been obtained using the calculated geometries. Our structural and electronic results at DFT level are comparable to data that can be found in literature (see following section).

We apply the standard $G_0W_0$ calculations to obtain the quasi-particle corrections to the energy levels, starting from the DFT eigenvalues and eigenfunctions[78, 89], as the electronic gap is underestimated in DFT[78-79], and also the relative positions of *s-p-d* levels can be affected by this description,

Quasi-particle energies are calculated as corrections to Kohn-Sham eigenvalues, and the proper inclusion of exchange and correlation produces, as shown in the following, an electronic gap in good agreement with experiments. The screening in the $G_0W_0$ calculation is treated within the plasmon pole approximation. The effect of the so-called local field effects in the screening calculation has been taken into account and carefully converged.

When going from one-electron excited properties, such as the ones described by the electronic levels, to two-particles properties, like the absorption of light, a further step in the many body description is necessary, namely, to include the description of the interaction between the excited electron and the hole left in the valence region. As shown in the following, the calculation of the optical absorption spectrum as a sum over transitions of independent particles (RPA)[85], gives in general an incorrect shape and intensity of the peaks, while the inclusion of many body effects gives the proper optical gap, and a good description of the absorption spectrum and of excitons. The inclusion of the excitonic Hamiltonian[85] in the calculation of the optical response provides excitonic eigenvalues and excitonic eigenstates produced by the diagonalization of the excitonic Hamiltonian. The electronic levels are mixed to produce optical transitions, which are not anymore between pairs of independent particles. The resulting eigenvalues describe the excitonic energies, reported in the following. The optical spectra have been obtained at three different theoretical levels: i) RPA calculations based on the PBE energies (RPA@PBE); ii) RPA calculations based on the $G_0W_0$ corrected energies (RPA@GW); and iii) BSE calculations based on the $G_0W_0$ corrected energies (BSE@GW) and again on the screening obtained at the level i). Driven by the case of the optical spectrum of rutile for the light polarization perpendicular to c direction, the $G_0W_0$ effects on the screening was also taken into account at the level iii) ($BSE_2$). The reason for this is detailed in the next section. The $G_0W_0$, optical RPA and BSE calculations have been performed with the code Yambo.[90]

## 3. Results and discussion

### 3.1 Electronic Properties

We calculated, at the DFT level, for both rutile and anatase, the band structure (Figs. 2-3) along the high symmetry directions, the density of states, and analyzed the spatial behaviour of wavefunctions involved in relevant bonds of the system. We do not report here these results, because they are in line with data already published.[54, 73, 75, 83, 91] We can say that the comparison with experimental photoemission spectra, integrated or angular resolved, is quite good, in spite of the putative difficulties of DFT in describing the *d*-electron properties.[70, 81, 92]

To take into account the importance of a different wavefunction starting point for the $G_0W_0$ calculations onto the electronic description of $TiO_2$, we compare, for the bulk material, the spatial localization of Kohn-Sham wavefuntions obtained from DFT-PBE[93] and from DFT-PBE0.[94-95] In PBE0, the parametrized description of exchange and correlation contributions is expected to improve the description of electron localized systems. The results in Fig. 4 indicate that the spatial differences between PBE and PBE0 wavefuntions are negligible. As it is expected, PBE wavefunctions show slightly more covalence in the bond between the 2*p* orbitals of the $O^{2-}$ ions and the 3*d* orbitals of the $Ti^{4+}$ ion. This effect is more noticeable in the valence band, but in any case the difference in the covalence is below 1% (the valence band has a 14% weight on 2*p* functions of $O^{2-}$ in PBE results and 13% in PBE0 ones). It can be inferred therefore that the election of the functional is not crucial to carry out the $G_0W_0$ calculations for the defect-free $TiO_2$ bulk, since both wavefunctions are very similar. Henceforth all the $G_0W_0$ results showed in this work will have as starting point the PBE wavefunctions. Our many body description provides results in quite good agreement with experimental data. As, however, the problem of initial wavefunctions, and *d*-electrons localization description, arises for defective systems with oxygen vacancies,[76, 96-99] we cannot exclude a major role of exchange-correlation on starting wavefunctions, and therefore on electronic-optical properties, when defective systems are considered.[72]

The electronic gap, corresponding to the difference between the VBM and the CBM, is calculated (PBE) as 1.93 eV for rutile (direct gap) and 2.15 eV for anatase (indirect gap), respectively (Table 1). The direct gap of anatase is 2.43 eV. The electronic gaps for the two phases are underestimated by almost 1.5-2.0 eV with respect to the (few) experimental data available. Indeed the smallest measured electronic gap for rutile is 3.3±0.5 eV.[56] The application of $G_0W_0$ to the two systems gives a correction of 1.66 eV and 1.81 eV for rutile and anatase, respectively, calculated at the respective k points of interest. The resulting electronic gaps (direct gap of 3.59 eV for rutile, indirect gap of 3.96 eV and direct one of 4.24, for anatase) are therefore in quite good agreement with the range of GW theoretical values (3.7-4.2 eV) reported in literature.[69-70, 73,82] The electronic gap for rutile is in good agreement with the experimental estimation given by various techniques, in particular with the direct estimation (3.3÷3.6 eV) given by the combined photoemission and inverse photoemission experiments.[56, 65] The agreement is worse with respect to the value of 4 eV obtained by a different photoemission experiment.[54]

For both phases, as expected, the electronic gap is definitely larger than the experimental optical gap of 3-3.2 eV obtained by, e.g., absorption experiments.[57,34]

In Figs. 2-3 the $G_0W_0$ corrections for high symmetry points of the BZ for rutile and anatase are also reported, plotted as dots ontop of the DFT band structure. While the corrections to the valence levels show a linear dependence with the distance from the gap, the corrections to the conduction levels have a more complex behaviour, due to the *d*-nature of those states. The effect of the nonlinear dependence of the quasi-particle corrections had to be evaluated in the optical spectrum, as discussed in the following paragraph. From Figs. 2-3, it is in any case clear that the proper inclusion of correlation through many-body treatment does not change in a dramatic way the relative positions of *d*-bands. Even if the use of a rigid scissor operator has to be carefully evaluated for this material, we will show that its use is reasonable in the optical calculations, instead of the full quasi-particle calculation over the IBZ.

### 3.2 Optical Absorption Properties

The optical absorption spectra calculated for the two phases, with polarization along the directions *x* (or the equivalent one *y*, in the following also called in-plane) and *z* (out-of-plane, along the c axis of the cells) of the unitary cells, are reported in Figs. 5 and 6. The spectra given by independent-particle transitions (RPA@PBE level) present two characteristic features: the band edge is underestimated due to the electronic gap underestimation in DFT, and the overall shape of the spectrum is, for both phases, and both orientations, different from the experiment, in the sense that the oscillator strengths are not correct. The inclusion of the quasi-particle description (RPA@GW), which improves the electronic gap description, does not affect the overall shape of the spectrum, and acts effectively as a scissor operator. We checked that the spectrum shape is almost unchanged, at RPA@GW level, if the two methods (full ab-initio calculation of $G_0W_0$ corrections for all the eigenvalues involved in the transitions, or ab-initio calculation of $G_0W_0$ correction just for the electronic gap) are used (spectra not shown). The absorption edge of RPA@GW is shifted to higher energies, with respect to RPA@PBE, even higher than expected from absorption experiments. The description of optical properties within the two interacting quasi-particles scheme by solving the BSE gives indeed a definite improvement of the result (BSE@GW). There is a significant agreement in the spectrum shape with respect to experiments, indicating that both electronic band gap and direct transitions are properly described by many body methods. The shape of the spectrum is now well described, with a redistribution of transitions at lower energies, with respect to the RPA results. This weigth redistribution of oscillator strengths is due to the inclusion of electron-hole interaction in the optical properties. The calculated optical onset is discussed in the following subsection, based on optical transitions evaluated from BSE eigenvalues calculations.

Going into a more detailed discussion, the agreement for spectral shape is in general very good for both polarizations in the case of anatase, as shown in Fig. 6. The optical absorption edge is well reproduced along both *xy* and *z*. In the in-plane *xy* direction, we obtain the initial sharp peaks whose shape, with respect to the RPA calculation, clearly indicates an excitonic contribution. In the *z* direction, the initial peak is however too much intense with respect to the experimental data. A possible over-estimation of excitonic effect for this peak is discussed at the end of this section (see also insets of Fig. 6).

The correspondence between experiment and our calculated spectrum is also quite good in the case of the *z*-polarization for rutile. The absorption shape is quite well reproduced, the sharp edge of the onset of direct transitions is correct, and also the relative intensities are in good agreement with the optical data (but the intensity of the first peak around 4 eV is underestimated, and this is not corrected by changing the screening description, see insets of Fig. 5 and the end of the section). The *xy* polarization in rutile seems to present some more difficulties in the absorption edge determination. Indeed, the overall spectrum shape is quite good, due to the exact description of excitonic effects and electron-hole exchange through BSE solution. The whole spectrum seems however to be shifted by about 0.1-0.2 eV to higher energy with respect to experiments.

Much better agreement can be obtained for rutile in-plane polarization if, when calculating the screening used in the BSE, the opening in the gap produced by $G_0W_0$ corrections is taken into account (see inset of Fig.5, left, $BSE_2$). A larger electronic gap makes the screening less efficient. Therefore the electron-hole interaction is less screened and the interacting levels become more bound. The inclusion of the same effect in the absorption along *z* does not affect the spectrum (see inset of Fig.5, right), because the description of the screening along the c axis of rutile is already adequate at RPA@PBE level. These differences in the results are related to the large anisotropy in the electronic dielectric constant of rutile (it is 8.427 parallel to c axis and smaller, 6.843, perpendicular to c axis[100]). For anatase, the electronic dielectric constants (5.41 and 5.82 parallel and perpendicular to c axis, respectively) are much more similar[101], and this results in a comparable description of the BSE optical properties based on the RPA screening. For anatase, the use of the $G_0W_0$ gap in the BSE screening improves slightly the edge of the absorption, but the overall intensities are in worse agreement with experiments (insets of Fig. 6). Both the first excitonic peak in *xy* and the first very intense peak in *z* are too much intense in this case.

### 3.3 Analysis of Excitonic Transitions

As the calculated absorption spectra by many body methods give in general a good agreement with experimental data, one should expect a reasonably good description of excitonic transitions energies. The analysis of energy, k-points and wavefuntions characterizing optical transitions, as obtained by diagonalizing the Bethe-Salpeter Hamiltonian, are described in the following and reported in Table 2. We explain the properties of excitons in the rutile and anatase phases, in both real and reciprocal space, and how their localization affects the optical absorption spectrum. The analysis of the spatial distribution of some excitons in the two phases is reported in Fig. 7, and provides very interesting information, when combined with the exciton binding energy data.

A direct comparison with data extrapolated from experiments, however, cannot be performed, because of the co-presence of several physical effects which are visible in experiments and not taken into account in our description. It is interesting to clarify the various contributions present in the experimental absorption spectra of this material, particularly when comparing with theoretical results.

The characteristics of excitons have been debated in this material. The experimental binding energy of the first, indirect exciton in rutile, at 3.031 eV, is reported to be 4 meV.[45, 57] Some uncertainty exists in the exact determination of optical edge. Moreover, the exciton is localized in anatase, and delocalized in rutile[34], at least based on experimental result, but an explanation of this behaviour is missing so far.

In $TiO_2$ polymorphs, direct optical transitions should appear at Γ for rutile and anatase (whose electronic gap is indirect, between Γ and a point close to X ). Due to symmetry of *p-d* wavefunctions, however, these transitions are symmetry-forbidden[31], and the first optical dipole transitions may occur involving other bands than the highest occupied and the lowest empty, or different k points of the BZ, as shown in the following.

At energies below the direct transition edge, other indirect transitions that occur through mediation by LO and TA phonons have been observed. The electron-phonon interaction is not included in the ab initio calculation of the direct transitions. Also the ionic screening, quite large in $TiO_2$[31], is not included. The polaron presence can further change the absorption properties, because of the local relaxation induced by the exciton creation. The reorganization energy associated to polarons in rutile and anatase is indeed larger than 1 eV, even if structural deformations are small.[32] The excitons here described are not interacting with the system polarized as a whole, but are screened just by the electronic component, and are not interacting with the phonons and the ionic relaxation induced by the exciton creation itself. This explains the differences between our results for optical transition energies and the values reported in literature. Moreover, defects such as oxygen vacancies, very frequent in this oxide, can easily affect the optical spectrum,[102] and defects are not taken into account in our results.

For rutile, threshold absorption measurements[45, 57] give a weak absorption peak at 3.031 eV, in the forbidden band gap. The attribution of this transition was not clear. It was first supposed to be due to a 1s exciton, allowed by a quadrupolar transition, and with a binding energy of 4 meV. The phonon contribution in this transition was suggested by Hyper-Raman scattering results.[52] Finally the peak was interpreted, in photoluminescence measurements, as a $2p_{xy}$ dipole-allowed exciton[57, 103], for light polarization parallel to *xy*. The experimental direct allowed gap (that is, the first intense structure in the absorption spectrum) is at 4.2 eV [57, 103]. In our ab initio treatment, processes involving electron-phonon coupling, or quadripolar transitions, are not taken into account. We obtain a good estimation of the intense optical edge in rutile, however. We calculate, for *xy* polarization, the most intense optically allowed transition at 4.24 eV (4.27 eV for *z* polarization), for k points belonging to the R-Z direction (R-X for *z* polarization) (see also Fig. 2). The states contributing to these transitions are the VBM and the CBM, and also transitions from VBM to CBM+4 are present. The transition at 4.24 eV corresponds to the first intense peak observed in the absorption spectra. However, we obtain other optical transitions, due to excitons, with oscillator strengths not negligible at energies below 4.24 eV.

The first three excitonic excitations here analyzed are given by transitions from the highest occupied *p*-band to the lowest empty *d*-band. The first two excitonic transitions (Fig. 7a) are optically symmetry-forbidden, for in-plane (*xy*) polarization, while the third transitions (Fig. 7b)) is optically active. It means that, for the first and second transitions, at 3.41 eV and 3.55 eV, the oscillator strengths have a quite low intensity, as they can be called dark excitons. The BZ point contributing to those transitions is mainly Γ, with other points outside of the *xy* plane. The third transition, at 3.59 eV, with larger oscillator strength, is the first bright excitons, and it has contributions from k points outside of *xy* plane, but it is forbidden at Γ for symmetry reasons. The energetic positions of these first transitions are 3.32, 3.51, 3.55 eV, when calculated with the screening modified by the $G_0W_0$ correction of energy levels (see insets of Fig. 5).

As already explained, these excitons are calculated without taking into account the effects of phonons, and therefore, appear at almost 0.4 eV higher energies than the observed forbidden optical gap. Transverse and optical phonons determine the lower optical edge observed in experiments.[34, 45, 57, 104] Direct transitions have been observed[104] at 3.57 eV for the in-plane directions, in quite good agreement with our estimation for the first bright exciton (3.59 eV).

With out-of-plane polarization, the first optically allowed transition is at 3.93 eV (the only experimental data to compare with is that direct transitions were observed above 3.6 eV[104] for polarization along the *z* direction), followed by other transitions at 4.03 and 4.06 eV. Transitions from the VBM to the CBM, CBM+1,+2 contribute, to these excitons, from various high symmetry k points distributed in all the BZ. The most intense transition is obtained at 4.27 eV, with contributions from k points on the R-X direction for the transition VBM-3 to CBM, and the direct transition at Γ from VBM to CBM.

The same transitions which are symmetry-allowed along *z* are optically forbidden with the in-plane polarization. Correspondingly, all other transitions, apart the ones here listed, are optically dark along the *z* axis.

The transitions plotted in Fig. 7 a-b are from O 2*p* states to Ti 3*d* states of the triply degenerate $t_{2g}$, as expected. While the first two dark exciton transitions involve Ti atoms farther away from the excited O atom, the optical active transition involves states of the nearest neighbour Ti atoms. The excitons are delocalized in a planar region, extending in the *z* direction. The bright optical exciton is localized (spread over three lattice constants) in the (110) direction (Fig 7c). This anisotropic distribution is also related to the anisotropic screening in rutile.

Concerning anatase, the fine structure of the absorption edge has been studied[34] at low temperature. The absorption edge, estimated at 3.2 eV[37], has a strong phonon contribution, as evident from the existence of an Urbach tail.[105-106] The phenomenon of Urbach tail has received different theoretical explanations[34], including exciton ionization, or exciton self-trapping, both of which require the participation of phonons. In the case of $TiO_2$, it has been attributed to exciton self-trapping[34,36,107], i.e. the localization of exciton through phonon coupling. It is favoured to occur in anatase with respect to rutile, because of its lower octahedral coordination. The Urbach Tail extends to energies below 3.4 eV

Our results for anatase show that, with in-plane polarization (*xy*), the first calculated transition is direct and the most intense one. It appears at 4.03 eV (3.93 with corrected screening, see insets of Fig. 6), and it is given by transitions at the k points along the Γ-Z direction of BZ. However the Γ point does not contribute to this exciton, for symmetry reasons. The following allowed transitions are at 4.13, 4.15, 4.23, 4.37 eV. (4.09, 4.12, 4.19, 4.30 eV with corrected screening), always along the Γ-Z direction, and with the two highest occupied and the two lowest empty states participating. The transitions at 4.15 and 4.23 eV also present a contribution from the Γ point. The transitions at 4.03 and 4.37 eV correspond, in the absorption spectrum, to the first and second sharp peaks observed in the experiments. Along *z* (out-of-plane) all the initial transitions allowed in the in-plane directions, are instead dark (not allowed for symmetry reasons) and the first optically allowed one is at 4.48 eV (4.34 with corrected screening), followed by a direct one at 4.53 eV (4.45 eV)). These optical transitions produce the intense peak reported in the spectrum around 4.5 eV. The contributing k-points lie along the Γ-Z direction. Involved states are again the two highest levels of the valence band, and the two lowest levels of the conduction band.

The spatial distribution (Fig. 7c) shows how the optical excitations, in anatase, have a strong localized character, when compared with rutile. The first exciton is extended for several (at least eight) lattice constants in the *xy* plane, and, following the chain-like structure of the crystal, is confined in the *z* direction (it extends along *z* about for one lattice constant, so it is almost confined in a single atomic plane).

To resume, the direct optical gaps, as estimated from BSE calculations, are in reasonable agreement with experimental data on direct transitions. For rutile, the intense direct optical edge of 4.24 eV is quite close to the electronic theoretical $G_0W_0$ gap we calculated (see Table 1), denoting a small exciton binding energy. For anatase, the optical gap is close but slightly larger than the indirect $G_0W_0$ gap, because given by direct transitions along the Γ-Z direction. It results to be smaller than the indirect gap when the $G_0W_0$ corrected screening is used, but the result in in the error bar for these calculations. In rutile we obtain indeed some excitons at energies lower than the direct transition at 4.24 eV, even if their intensities are quite low. In any case, a direct exciton (without interaction with phonons) has been observed[104] with an energy in quite good agreement with our result, at 3.57 eV. If we assume the value of 4.24 eV as reference energy, we obtain indeed a quite large binding energy for this low intensity exciton, of 0.65 eV. In anatase, the effect of electron-hole interaction is mainly visible in the intensity and in the shape of the spectrum, more than in the exciton binding energy.

Excitons in rutile and anatase have spatial distributions that are determined by the crystal structure and the screening properties of the materials. In both polymorphs, the excitonic wavefunctions are two-dimensional[108], and this planar distribution is localized in just one atomic layer in the anatase case. This stronger localization probably supports the possibility of experimentally reported self-trapping[34,36,107]. The behaviour of excitons in anatase have indeed also been modelled within a Franck-Condon model including 2D-exciton transitions.[108] For rutile, the more delocalized distribution of the excitonic wavefucntions can be connected with the observed[109] low electron-hole recombination rates in this phase.

The finding that excitons are strongly bidimensional in the (001) plane of anatase and almost bidimensional in the (110) plane of rutile, opens interesting scenarios concerning the control of the optical properties of nanoctructures with these external surfaces.

## 4. Conclusions

In this paper, we provided a theoretically consistent description of the properties of the two main crystalline phases of titania by means of first principle calculations based on DFT and GW-BSE. Apart from the advantage of describing, at the same level of theory, different properties of the two systems, we were able to get results in quite good agreement with available photoemission and optical experiments for both rutile and anatase. In particular, we showed that the electronic properties are well described when electron correlations are included in the calculation (at the $G_0W_0$ self-energy level starting from the GGA wavefunctions). Moreover, the optical absorption spectra are quantitatively in good agreement with experiments once electron-hole effects are included, even if still some minor discrepancies in the intensities are present for the spectra along different polarisation directions. The anisotropy of the material (in particular in the rutile phase) is shown to be important both in the optical spectra, and in the estimation of the screened interaction between hole and electron.

We gave here the first theoretical description of direct excitons in $TiO_2$, by estimating the first direct optical transition, and, with respect to it, the exciton binding energy. The excitonic effects act on optical spectra on redistributing the transitions weights, more than on changing the excitonic energies, especially in anatase. The spatial behaviour of optical excitations is found to be strongly anisotropic. As experimental results suggested, we found a stronger excitonic localization in anatase than in rutile. Excitons have, in both phases, a two-dimensional character, induced by the crystalline structure. In anatase the first bright exciton is completely localized in the (001) plane. In rutile the localization acts in the plane (110).

The present characterization of excitonic properties of rutile and anatase will help to understand later on the photophysical response of $TiO_2$ surfaces, nanowires, nanoclusters. The excitonic description of pure $TiO_2$ phases is also the reference starting point for investigation of exciton-phonon interactions, polarons, and defects effects on optical properties. Further work along these lines is now under progress.

**Acknowledgments**

We acknowledge funding by "Grupos Consolidados UPV/EHU del Gobierno Vasco" (IT-319-07), the European Community through e-I3 ETSF project (Contract Number 211956), ACI-Promociona (ACI2009-1036), the Army Research Office Grant No.

W911NF-07-1-0052, and National Science Foundation Grant No. CHE-0650756. We acknowledge support by the Barcelona Supercomputing Center, "Red Española de Supercomputacion", SGIker ARINA (UPV/EHU), Transnational Access Programme HPC-Europe++, and the supercomputing center of the Environmental Molecular Sciences Laboratory at PNNL, sponsored by the DOE Office of Biological and Environmental Research.

J.M. G-L. acknowledges funding Spanish MEC through 'Juan de la Cierva' program. L.C. acknowledges funding from UPV/EHU through the 'Ayudas de Especialización para Investigadores Doctores' program. H.P. thanks Ikerbasque for the support of his stay at DIPC Donostia International Physics Center. We aknowledge P. Giannozzi, M. Gatti and M. Palummo for fruitful discussions.

**References**

[1] A. Fujishima and K. Honda, Nature **238**, 37 (1972).
[2] B. Oregan and M. Gratzel, Nature **353**, 737 (1991).
[3] V. P. Indrakanti, J. D. Kubicki, and H. H. Schobert, Energy & Environmental Science **2**, 14 (2009).
[4] A. Fujishima, X. T. Zhang, and D. A. Tryk, Surface Science Reports **63**, 515 (2008).
[5] M. Gratzel, Nature **414**, 338 (2001).
[6] A. Hagfeldt and M. Gratzel, Accounts of Chemical Research **33**, 269 (2000).
[7] M. R. Hoffmann, S. T. Martin, W. Y. Choi, and D. W. Bahnemann, Chemical Reviews **95**, 69 (1995).
[8] S. U. M. Khan, M. Al-Shahry, and W. B. Ingler, Science **297**, 2243 (2002).
[9] U. Aschauer, Y. B. He, H. Z. Cheng, S. C. Li, U. Diebold, and A. Selloni, Journal of Physical Chemistry C **114**, 1278 (2010).
[10] X. Q. Gong, A. Selloni, M. Batzill, and U. Diebold, Nature Materials **5**, 665 (2006).
[11] X. Q. Gong, A. Selloni, O. Dulub, P. Jacobson, and U. Diebold, Journal of the American Chemical Society **130**, 370 (2008).
[12] Y. B. He, A. Tilocca, O. Dulub, A. Selloni, and U. Diebold, Nature Materials **8**, 585 (2009).
[13] M. Lazzeri and A. Selloni, Physical Review Letters **87**, 266105 (2001).
[14] S. C. Li, J. G. Wang, P. Jacobson, X. Q. Gong, A. Selloni, and U. Diebold, Journal of the American Chemical Society **131**, 980 (2009).
[15] A. Selloni, Nature Materials **7**, 613 (2008).
[16] A. Vittadini, M. Casarin, and A. Selloni, Theoretical Chemistry Accounts **117**, 663 (2007).
[17] A. Vittadini, A. Selloni, F. P. Rotzinger, and M. Gratzel, Physical Review Letters **81**, 2954 (1998).
[18] O. Bikondoa, C. L. Pang, R. Ithnin, C. A. Muryn, H. Onishi, and G. Thornton, Nat Mater **5**, 189 (2006).
[19] A. L. Linsebigler, G. Q. Lu, and J. T. Yates, Chemical Reviews **95**, 735 (1995).

20  A. Beltran, J. R. Sambrano, M. Calatayud, F. R. Sensato, and J. Andres, Surface Science **490**, 116 (2001).
21  R. Buonsanti, et al., Journal of the American Chemical Society **128**, 16953 (2006).
22  A. Iacomino, G. Cantele, D. Ninno, I. Marri, and S. Ossicini, Physical Review B **78**, 075405 (2008).
23  A. Petrella, M. Tamborra, M. L. Curri, P. Cosma, M. Striccoli, P. D. Cozzoli, and A. Agostiano, Journal of Physical Chemistry B **109**, 1554 (2005).
24  A. Kongkanand, K. Tvrdy, K. Takechi, M. Kuno, and P. V. Kamat, Journal of the American Chemical Society **130**, 4007 (2008).
25  Y.-L. Lee, B.-M. Huang, and H.-T. Chien, Chemistry of Materials **20**, 6903 (2008).
26  Y.-Y. Lin, T.-H. Chu, S.-S. Li, C.-H. Chuang, C.-H. Chang, W.-F. Su, C.-P. Chang, M.-W. Chu, and C.-W. Chen, Journal of the American Chemical Society **131**, 3644 (2009).
27  O. Niitsoo, S. K. Sarkar, C. Pejoux, S. Rühle, D. Cahen, and G. Hodes, Journal of Photochemistry and Photobiology A: Chemistry **181**, 306 (2006).
28  R. Plass, S. Pelet, J. Krueger, M. Gratzel, and U. Bach, The Journal of Physical Chemistry B **106**, 7578 (2002).
29  P. Yu, K. Zhu, A. G. Norman, S. Ferrere, A. J. Frank, and A. J. Nozik, The Journal of Physical Chemistry B **110**, 25451 (2006).
30  A. Zaban, O. I. Mićić, B. A. Gregg, and A. J. Nozik, Langmuir **14**, 3153 (1998).
31  C. Persson and A. F. da Silva, Applied Physics Letters **86**, 231912 (2005).
32  N. A. Deskins and M. Dupuis, Physical Review B **75**, 195212 (2007).
33  B. J. Morgan and G. W. Watson, Physical Review B **80**, 233102 (2009).
34  H. Tang, F. Levy, H. Berger, and P. E. Schmid, Physical Review B **52**, 7771 (1995).
35  H. Tang, K. Prasad, R. Sanjines, P. E. Schmid, and F. Levy, Journal of Applied Physics **75**, 2042 (1994).
36  K. Watanabe and T. Hayashi, Journal of Luminescence **112**, 88 (2005).
37  M. Cardona and G. Harbeke, Physical Review **137**, 1467 (1965).
38  D. C. Cronemeyer, Physical Review **87**, 876 (1952).
39  R. G. Breckenridge and W. R. Hosler, Physical Review **91**, 793 (1953).
40  F. Arntz and Y. Yacoby, Physical Review Letters **17**, 857 (1966).
41  A. Frova, P. J. Boddy, and Y. S. Chen, Physical Review **157**, 700 (1967).
42  A. K. Ghosh, F. G. Wakim, and R. R. Addiss, Physical Review **184**, 979 (1969).
43  D. W. Fischer, Physical Review B **5**, 4219 (1972).
44  S. Hufner and G. K. Wertheim, Physical Review B **7**, 2333 (1973).
45  J. Pascual, J. Camassel, and H. Mathieu, Physical Review B **18**, 5606 (1978).
46  L. A. Grunes, Physical Review B **27**, 2111 (1983).
47  R. G. Egdell, S. Eriksen, and W. R. Flavell, Solid State Communications **60**, 835 (1986).
48  F. M. F. Degroot, M. Grioni, J. C. Fuggle, J. Ghijsen, G. A. Sawatzky, and H. Petersen, Physical Review B **40**, 5715 (1989).
49  R. L. Kurtz, R. Stockbauer, T. E. Madey, E. Roman, and J. L. Desegovia, Surface Science **218**, 178 (1989).
50  G. Vanderlaan, Physical Review B **41**, 12366 (1990).
51  R. Heise, R. Courths, and S. Witzel, Solid State Communications **84**, 599 (1992).
52  K. Watanabe, K. Inoue, and F. Minami, Physical Review B **46**, 2024 (1992).

53   F. M. F. Degroot, J. Faber, J. J. M. Michiels, M. T. Czyzyk, M. Abbate, and J. C. Fuggle, Physical Review B **48**, 2074 (1993).
54   P. J. Hardman, G. N. Raikar, C. A. Muryn, G. Vanderlaan, P. L. Wincott, G. Thornton, D. W. Bullett, and P. Dale, Physical Review B **49**, 7170 (1994).
55   R. Sanjines, H. Tang, H. Berger, F. Gozzo, G. Margaritondo, and F. Levy, Journal of Applied Physics **75**, 2945 (1994).
56   Y. Tezuka, S. Shin, T. Ishii, T. Ejima, S. Suzuki, and S. Sato, Journal of the Physical Society of Japan **63**, 347 (1994).
57   A. Amtout and R. Leonelli, Physical Review B **51**, 6842 (1995).
58   N. Hosaka, T. Sekiya, M. Fujisawa, C. Satoko, and S. Kurita, J. . Electron Spec. Related Phenomena **78**, 75 (1996).
59   N. Hosaka, T. Sekiya, C. Satoko, and S. Kurita, Journal of the Physical Society of Japan **66**, 877 (1997).
60   M. Arai, S. Kohiki, H. Yoshikawa, S. Fukushima, Y. Waseda, and M. Oku, Physical Review B **65**, 085101 (2002).
61   N. Vast, L. Reining, V. Olevano, P. Schattschneider, and B. Jouffrey, Physical Review Letters **88**, 037601 (2002).
62   A. G. Thomas, et al., Physical Review B **67**, 035110 (2003).
63   M. Launay, F. Boucher, and P. Moreau, Physical Review B **69**, 035101 (2004).
64   A. G. Thomas, et al., Physical Review B **75**, 035105 (2007).
65   S. Rangan, S. Katalinic, R. Thorpe, R. A. Bartynski, J. Rochford, and E. Galoppini, Journal of Physical Chemistry C, 1139 (2010).
66   L. Kavan, M. Gratzel, S. E. Gilbert, C. Klemenz, and H. J. Scheel, Journal of the American Chemical Society **118**, 6716 (1996).
67   M. Launay, F. Boucher, and P. Moreau, Physical Review B **69**, 035101 (2004).
68   N. Vast, L. Reining, V. Olevano, P. Schattschneider, and B. Jouffrey, Physical Review Letters **88**, 4 (2002).
69   M. van Schilfgaarde, T. Kotani, and S. Faleev, Physical Review Letters **96** (2006).
70   T. Kotani, M. van Schilfgaarde, S. V. Faleev, and A. Chantis, in *1st International Conference on Quantum Simulators and Design*, Hiroshima, JAPAN, 2006).
71   J. Muscat, A. Wander, and N. M. Harrison, Chemical Physics Letters **342**, 397 (2001).
72   A. Janotti, J. B. Varley, P. Rinke, N. Umezawa, G. Kresse, and C. G. Van de Walle, Physical Review B **81**, 085212 (2010).
73   M. Oshikiri, M. Boero, J. Ye, F. Aryasetiawan, and G. Kido, in *3rd International Symposium on Transparent Oxide Thin Films for Electronics and Optics (TOEO-3)* (Elsevier Science Sa, Tokyo, Japan, 2003), p. 168.
74   H. M. Lawler, J. J. Rehr, F. Vila, S. D. Dalosto, E. L. Shirley, and Z. H. Levine, Physical Review B **78**, 205108 (2008).
75   F. Labat, P. Baranek, C. Domain, C. Minot, and C. Adamo, Journal of Chemical Physics **126** (2007).
76   M. V. Ganduglia-Pirovano, A. Hofmann, and J. Sauer, Surface Science Reports **62**, 219 (2007).
77   U. Diebold, Surface Science Reports **48**, 53 (2003).
78   R. W. Godby, M. Schlüter, and L. J. Sham, Physical Review B **37**, 10159 (1988).
79   R. O. Jones and O. Gunnarsson, Reviews of Modern Physics **61**, 689 (1989).

80  M. Mikami, S. Nakamura, O. Kitao, H. Arakawa, and X. Gonze, Japanese Journal of Applied Physics Part 2-Letters **39**, L847 (2000).

81  M. van Schilfgaarde, T. Kotani, and S. Faleev, Physical Review Letters **96**, 226402 (2006).

82  L. Thulin and J. Guerra, Physical Review B **77** (2008).

83  R. Asahi, Y. Taga, W. Mannstadt, and A. J. Freeman, Physical Review B **61**, 7459 (2000).

84  The GGA in the PBE [J. P. Perdew, K. Burke and M. Ernzerhof, Physical Review Letters **77**, 3865-3868 (1996)] parameterization for the exchange-correlation potential was employed. Hybrid functional calculations have been performed with PBE0 [C. Adamo and V. Barone, Journal of Chemical Physics **110** (13), 6158-6170 (1999); M. Ernzerhof and G. E. Scuseria, Journal of Chemical Physics **110**, 5029-5036 (1999)]. Semicore $3s$ and $3p$ states have been included explicitly in the Ti pseudopotential, due to their noticeable exchange interaction with the valence states [L. Thulin and J. Guerra, Physical Review B **77**, 195112 (2008)]. Wavefunction cut-off of 170 Ryd was used to ensure convergence on structural properties. Convergence on energy cut-off has been carefully checked up to 0.01 eV on total energies and 0.001 eV on eigenvalues, using up to 512 k points (8×8×8 mesh in the Monkhorst-Pack scheme [H. J. Monkhorst and J. D. Pack, Physical Review B **13**, 5188-5192 (1976)] for both phases in the irreducible Brillouin Zone (IBZ). K-points meshes of 8×8×8 have been used for both the optical and $G_0W_0$ calculations.

85  G. Onida, L. Reining, and A. Rubio, Reviews of Modern Physics **74**, 601 (2002).

86  P. Hohenberg and W. Kohn, Physical Review **136**, B864 (1964).

87  W. Kohn and L. J. Sham, Physical Review **140**, A1133 (1965).

88  P. Giannozzi, et al., Phys.: Condens. Matt. **21**, 395502 (2009)

89  F. Aryasetiawan and O. Gunnarsson, Reports on Progress in Physics **61**, 237 (1998).

90  A. Marini, C. Hogan, M. Grüning, and D. Varsano, Computer Physics Communications **180**, 1392 (2009).

91  K. M. Glassford and J. R. Chelikowsky, Physical Review B **46**, 1284 (1992).

92  M. Gatti, F. Bruneval, V. Olevano, and L. Reining, Physical Review Letters **99**, 266402 (2007).

93  J. P. Perdew, K. Burke, and M. Ernzerhof, Physical Review Letters **77**, 3865 (1996).

94  C. Adamo and V. Barone, Journal of Chemical Physics **110**, 6158 (1999).

95  M. Ernzerhof and G. E. Scuseria, Journal of Chemical Physics **110**, 5029 (1999).

96  C. Di Valentin, G. Pacchioni, and A. Selloni, Physical Review Letters **97**, 166803 (2006).

97  E. Finazzi, C. Di Valentin, G. Pacchioni, and A. Selloni, Journal of Chemical Physics **129**, 154113 (2008).

98  P. Kruger, et al., Physical Review Letters **100**, 055501 (2008).

99  T. Minato, et al., Journal of Chemical Physics **130**, 124502 (2009).

100 T. A. Davis and K. Vedam, Journal of the Optical Society of America **58**, 1446 (1968).

101 R. J. Gonzalez, R. Zallen, and H. Berger, Physical Review B **55**, 7014 (1997).

102 T. Sekiya, S. Kamei, and S. Kurita, Journal of Luminescence **87-89** (2000).

103 A. Amtout and R. Leonelli, Physical Review B **46**, 15550 (1992).

[104] O. Madelung, U. Rössler, and M. e. Schulz, SpringerMaterials - The Landolt-Börnstein Database (http://www.springermaterials.com).
[105] M. Capizzi and A. Frova, Phys. Rev. Lett. **25**, 1298 (1970).
[106] F. Urbach, Phys. Rev. **92** (1953).
[107] N. Hosaka, T. Sekiya, and S. Kurita, Journal of Luminescence **72-74** (1997).
[108] T. Makino, Y. Segawa, M. Kawasaki, Y. Matsumoto, H. Koinuma, M. Murakami, and R. Takahashi, Journal of the Physical Society of Japan **72**, 2696 (2003).
[109] R. Katoh, M. Murai, and A. Furube, Chemical Physics Letters **461**, 238 (2008).
[110] J. K. Burdett, T. Hughbanks, G. J. Miller, J. W. Richardson, and J. V. Smith, Journal of the American Chemical Society **109**, 3639 (1987).

**Tables**

**Table 1.** Calculated electronic band gaps, in eV, for rutile and anatase at different levels of theory. Rutile direct gap is at Γ, but quite close levels can be found at M and along Γ-M. For anatase indirect and direct electronic gaps are reported. The indirect gap is between X and Γ, the direct at Γ. The optical band gaps are also shown for comparison purposes.

|  | PBE$_{previous\ works}$ | PBE$_{this\ work}$ | GW$_{previous\ works}$ | G$_0$W$_0$ $_{this\ work}$ | Direct+Inverse Photoemission | optical absorption edge (indirect) |
|---|---|---|---|---|---|---|
| Rutile | 1.88$^{75,\ 80}$ | **1.93** | 3.78$^{69-70}$ | **3.59** | 3.3, 3.6$^{56,\ 65}$ | 3.0$^{57}$ |
| Anatase | 2.05/2.36$^{75,\ 80}$ | **2.15/2.43** | 3.79$^{82}$ | **3.96/4.24** | --- | 3.2$^{34}$ |

**Table 2.** Description of optical transitions, including excitonic effects. In parenthesis, the values with the screening corected by the G$_0$W$_0$ gap. Experimental optical data are only available for rutile, but for anatase the indirect and direct G$_0$W$_0$ gaps (3.96 eV, 4.24 eV) can be used for reference. The VBM used in the text corresponds to band n. 24, the CBM to n.25.

|  | Energy (eV) | Experimental energy (eV) | k-points | Bands |
|---|---|---|---|---|
| Rutile xy | 3.59 (3.55) | 3.57$^{104}$ (exciton) | close to (001) plane | 24→25 |
|  | 4.24 (4.03) | 4.2 $^{57,\ 103}$ (direct optical gap) | R-Z | 24;→25,29 |
| Rutile z | 3.93, 4.03, 4.06 (3.89, 3.99, 4.02) | > 3.6 eV $^{104}$ | all BZ | 24→25-27 |
|  | 4.27 (4.23) | -- | R-X ; Γ | 21→25; 24→25 |
| Anatase xy | 4.03, 4.13 (3.93, 4.09) | -- | Γ-Z | 23,24→25,26 |
|  | 4.15, 4.23 (4.12, 4.19) | -- | Γ-Z; Γ | 23,24→25,26 |
|  | 4.37 (4.30) | -- | Γ-Z | 23,24→25,26 |
| Anatase z | 4.48 (4.34) | -- | Γ-Z | 23,24→25,26 |
|  | 4.53 (4.45) | -- | Γ-Z | 23,24→25,26 |

**Figure Captions**

**Figure 1.** (Color online) Tetrahedral unit cell (left), crystallographic structure (center) and $TiO_6$ octahedrons (right) of rutile (top); and tetrahedral conventional cell, crystallographic structure (center) and $TiO_6$ octahedrons (right) of anatase (bottom). Experimental lattice constants and Ti-O distances are showed (in Å). Ti-O distances are larger along the axial direction than in the planar direction of the octahedron. The two phases for rutile and anatase belong to the $P4_{mnm}$ and $I4_{amd}$ spatial groups, respectively[110]. Rutile has a tetragonal unit cell, with 6 atoms, whereas anatase has a body centered tetragonal conventional cell, with 6 atoms. The calculated lattice constants in this work are a = 4.65 Å, c = 2.97 Å for rutile and a=3.81 Å, c= 9.64 Å for anatase, very close to the experimental values (less than 3% error).

**Figure 2.** (Color online) Electronic band structure of rutile bulk, along the high symmetry directions of the First Brillouin Zone. Black lines indicate the GGA calculation, yellow points indicate the values obtained with the $G_0W_0$ corrections.

**Figure 3.** (Color online) Electronic band structure of anatase. bulk, along the high symmetry directions of the First Brillouin Zone. Black lines indicate the GGA calculation, yellow points indicate the values obtained with the $G_0W_0$ corrections.

**Figure 4.** (Color online) PBE and PBE0 wavefuntions (and the difference between them) for the valence and conduction bands at Γ point for the anatase phase. Contour plots (absolute values of the wavefunctions) of the plane z = 0 are also shown below their corresponding 3-D plots. Notice that the scale for the differences is ten times smaller that the one for PBE and PBE0 results. In the case of the valence band the difference in the covalence (labeled CD) is indicated with red arrows.

**Figure 5.** (Color online) Imaginary part of the dielectric constant for rutile, in-plane polarization xy (left) and out-of-plane polarization z (right), calculated by GGA Random Phase Approximation (RPA@PBE, dashed blue line) using $G_0W_0$ on top of GGA (RPA@GW, dotted light green), and *via* Bethe-Salpeter Equation (BSE@GW, black solid line). The experimental spectrum (solid dotted black, ref. [37]) is also shown

for comparison purposes. Insets: BSE spectrum (BSE$_2$, red solid line) calculated by including in the screening calculation the proper G$_0$W$_0$ electronic gap.

**Figure 6.** (Color online) Imaginary part of the dielectric constant for anatase, in-plane polarization xy (left) and out-of-plane polarization z (right), calculated by GGA Random Phase Approximation (RPA@PBE, dashed blue line) using G$_0$W$_0$ on top of GGA (RPA@GW, dotted light green), and *via* Bethe-Salpeter Equation (BSE@GW, black solid line). The experimental spectrum (solid dotted black, ref. [37]) is also shown for comparison purposes. Insets: BSE spectrum (BSE$_2$, red solid line) calculated by including in the screening calculation the proper G$_0$W$_0$ electronic gap.

**Figure 7.** (Color online) a) Spatial distribution (yellow isosurfaces in arbitrary units) of the first partial dark exciton at 3.41 eV in rutile, for in-plane polarization. The second exciton, at 3.55 eV, is dark, and shows a spatial wavefunction similar to the one of the first exciton. b) Spatial distribution (yellow isosurfaces in arbitrary units) of the third, optically active exciton in rutile, at 3.59 eV, for in-plane polarization. c) Spatial distribution (yellow isosurfaces in arbitrary units) of the first direct exciton in anatase, at 4.03 eV, in-plane polarization. The hole position in a)-c) is denoted by the light green dot, located on a O atom.

**Figure 1.**

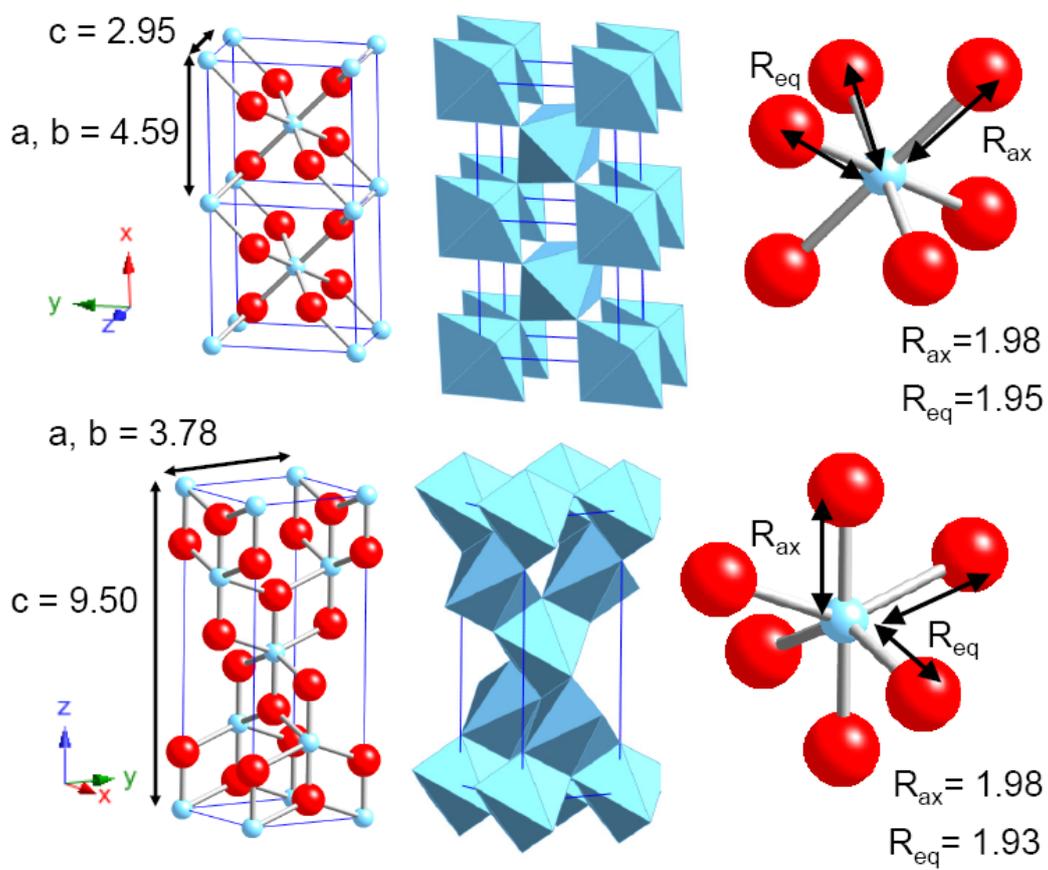

**Figure 2**

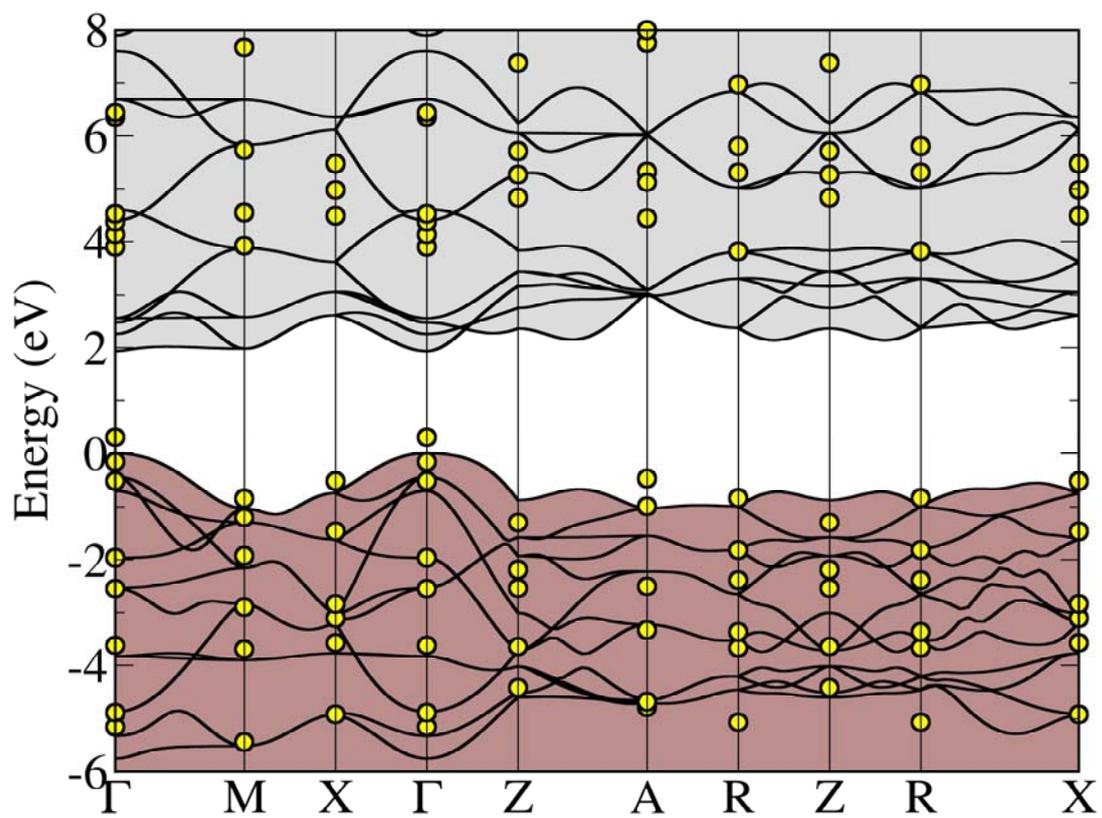

**Figure 3**

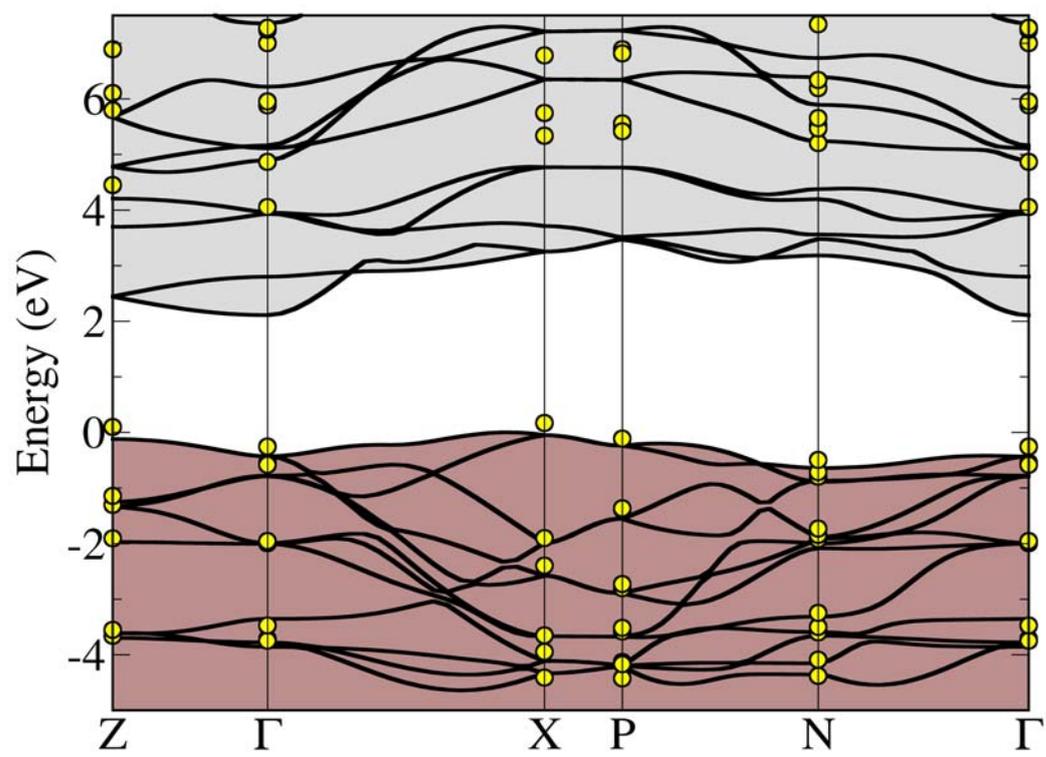

**Figure 4**

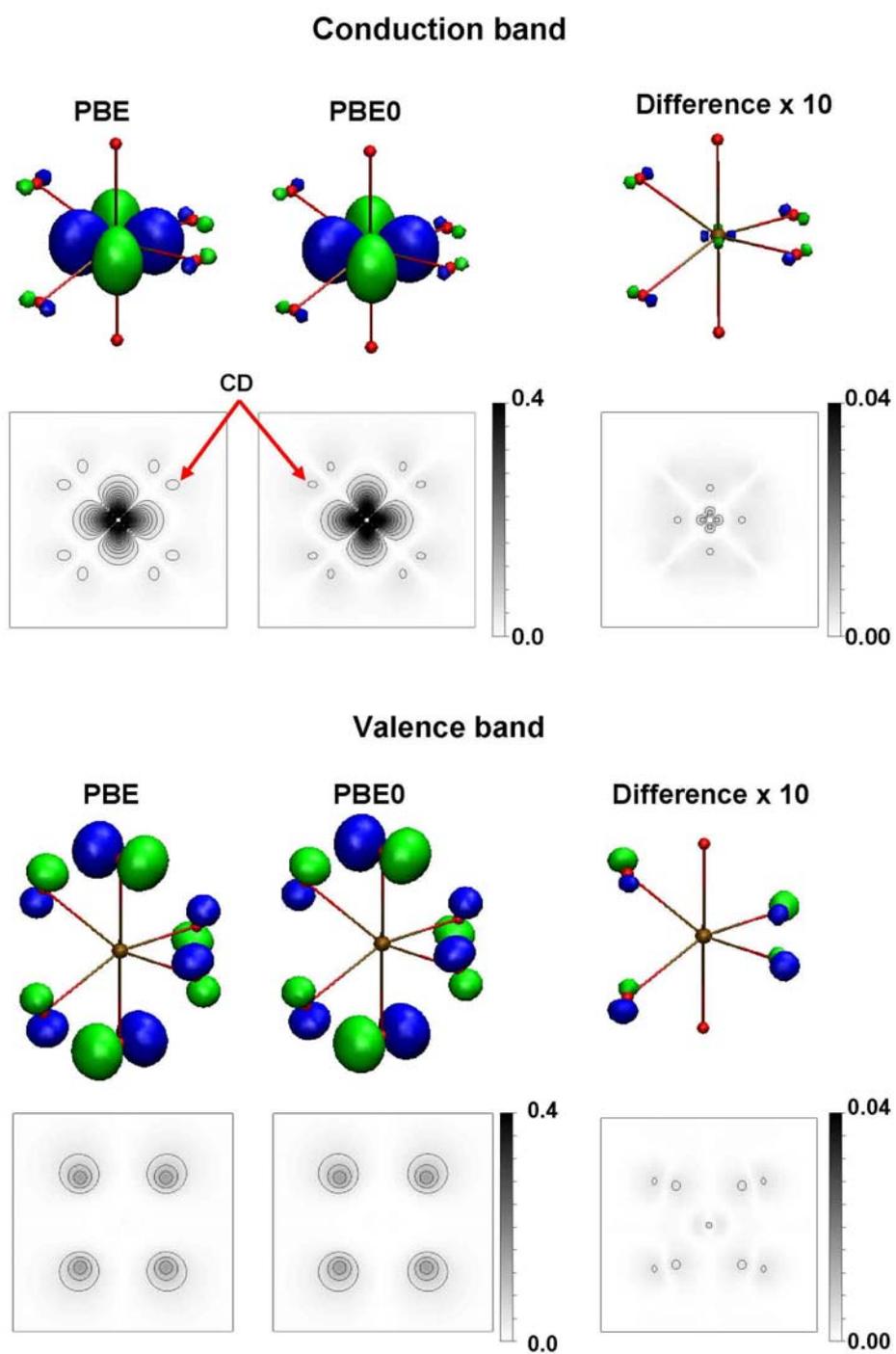

**Figure 5**

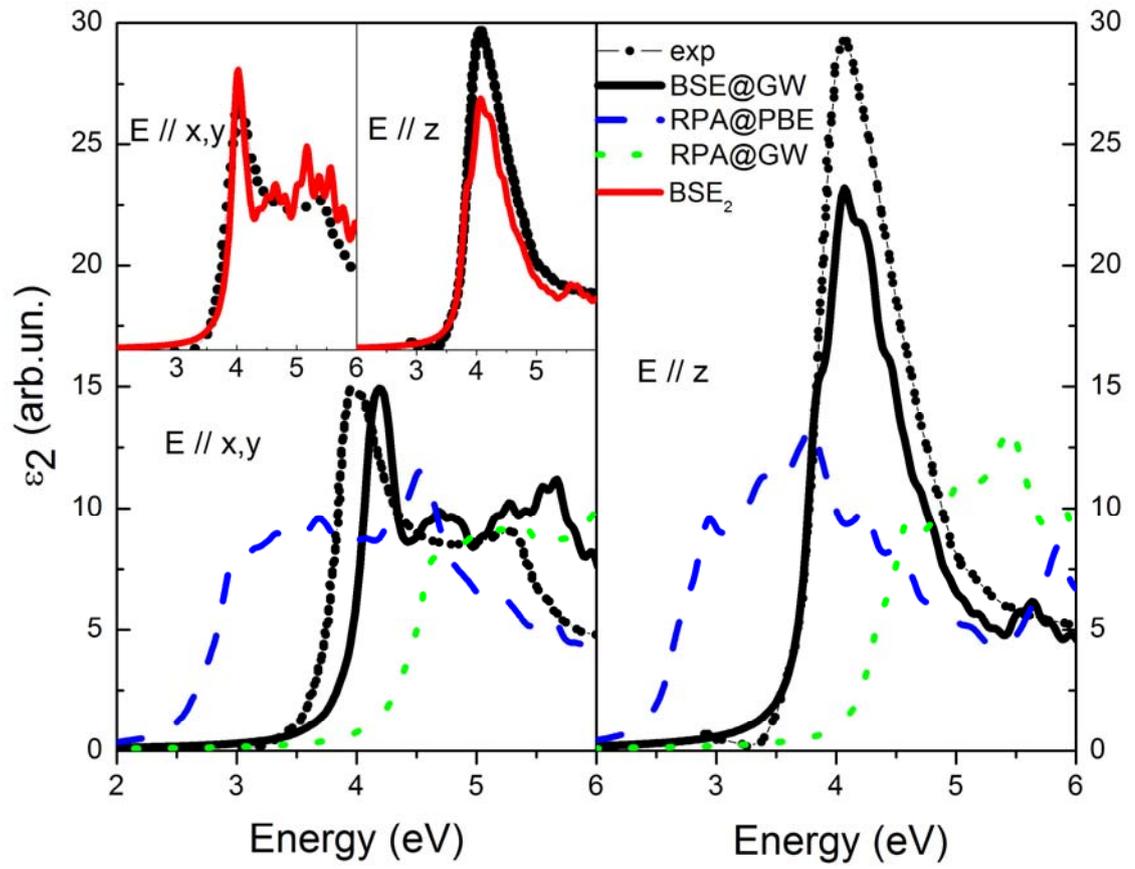

**Figure 6**

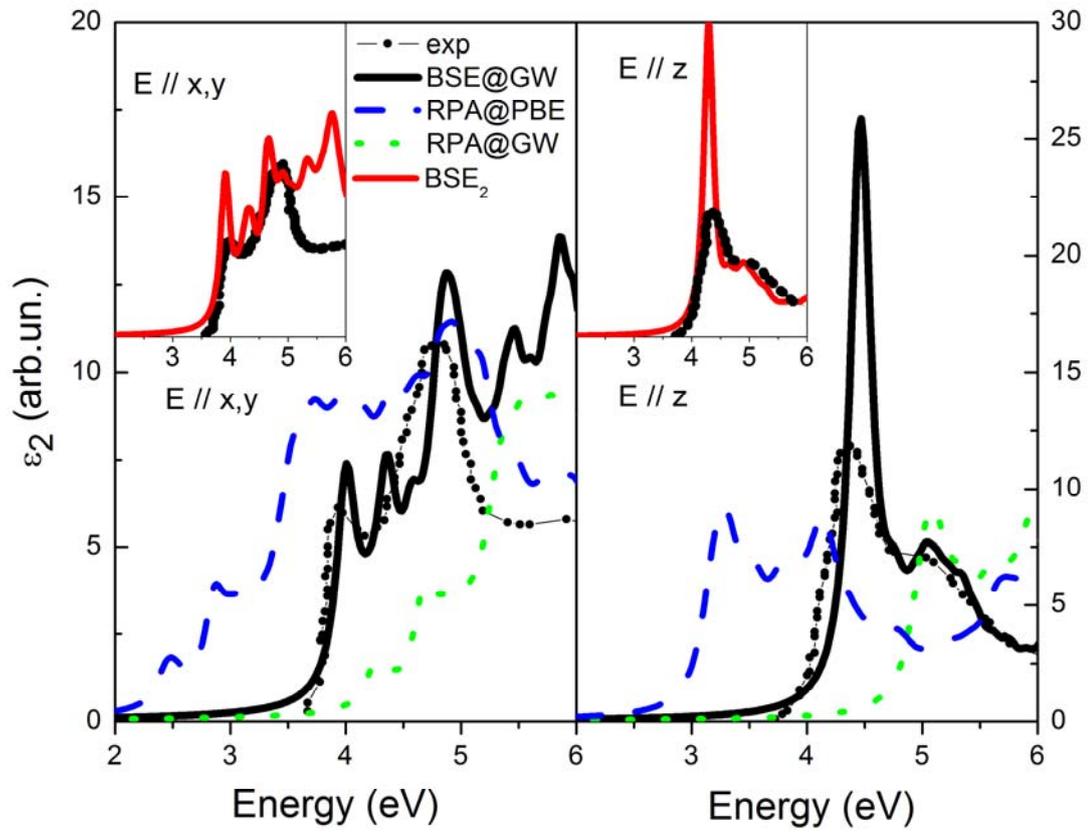

**Figure 7**

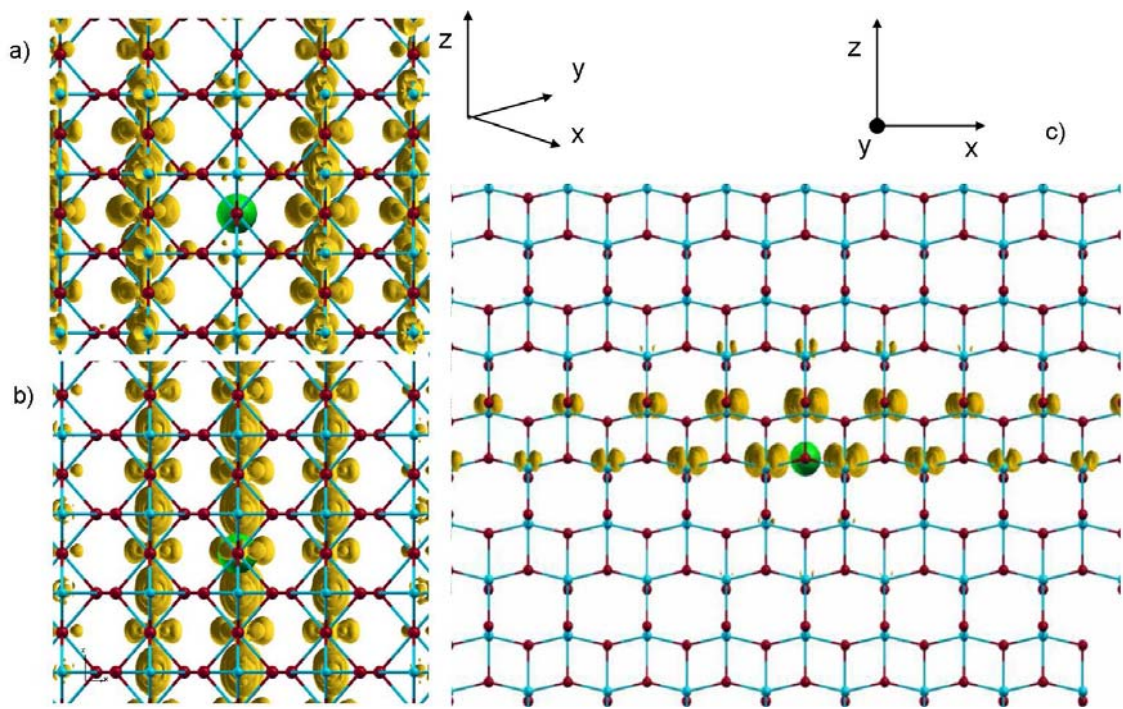